\documentclass[twocolumn,showpacs,preprintnumbers,amsmath,amssymb]{revtex4}
\usepackage{mathrsfs}

\usepackage{graphicx}
\usepackage{dcolumn}
\usepackage{bm}

\begin{document}

\preprint{}

\title{Non-Abelian Josephson effect between two spinor Bose-Einstein condensates in double optical traps}

\author{Ran Qi$^{1}$, Xiao-Lu Yu$^{1,2}$, Z. B. Li$^{2}$ and W. M. Liu$^{1}$}
\address{$^1$Beijing National Laboratory for Condensed Matter Physics,
Institute of Physics, Chinese Academy of Sciences, Beijing 100080,
China}
\address{$^2$The State Key Laboratory of Optoelectronic Materials and Technologies,
School of Physics and Engineering, Sun Yat-Sen University, Guangzhou 510275, China}

\date{\today}

\begin{abstract}
We investigate the non-Abelian Josephson effect in spinor
Bose-Einstein condensates with double optical traps. We propose, for
the first time, a real physical system which contains non-Abelian
Josephson effects. The collective modes of this weak coupling system
have very different density and spin tunneling characters comparing
to the Abelian case. We calculate the frequencies of the pseudo
Goldstone modes in different phases between two traps respectively,
which are a crucial feature of the non-Abelian Josephson effects. We
also give an experimental protocol to observe this novel effect in
future experiments.
\end{abstract}


\maketitle \emph{Introduction}.---The Josephson effect is a quantum
tunneling phenomena which occurs when a pair of superconducting or
superfluid systems are connected together by a weak link due to some
kind of physical barrier. Beginning with Josephson's original paper
in 1962 \cite{Josephson}, the Josephson effect has become a paradigm
of the phase coherence manifestation in a macroscopic quantum
system. With the amazing experimental progress in cold atom, the
Josephson junction has been realized for the trapped Bose-Einstein
condensates (BEC) of $^{87}Rb$ \cite{M. Albiez}. However, most of
extensive studies about this effect focus on the Abelian case  so
far, in terms of a junction of two systems with spontaneously broken
Abelian symmetry \cite{Smerzi,Barone,Chudnovsky,Gunther}.

Recently, F. Esposito \textit{et al.} generalized the Abelian
Josephson effect to the non-Abelian case in a field theoretic
language \cite{Esposito}. The non-Abelian Josephson effect emerges
in a junction of two systems with spontaneously broken non-Abelian
gauge symmetries, which are weakly coupled together. These two
systems should have the same symmetry so that there is an initial
doubling of symmetry. This doubled symmetry should be spontaneously
broken and leads to a doubling of the corresponding Goldstone modes.
However, due to the coupling between the two systems, only the
diagonal part of the symmetry group generators is preserved, which
will give rise to one set of true gapless Goldstone modes and one
set of pseudo Goldstone modes. These pseudo Goldstone modes is
gapped and leads to the non-Abelian Josephson effect. The
non-Abelian Josephson effect is an universal effect in physics and
it exists in a lot of physical system with a non-Abelian symmetry.
There are some other possible scenarios for constructing this
effect, for example, in the superfluid $^{3}He$ Joesphson weak link
\cite{Davis}, high density phases of QCD \cite{Rajagopal}, high Tc
superconductor system with SO(5) symmetry \cite{sczhang} and
nonlinear optics induced non-Abelian field \cite{Ruseckas}. However,
there are no any specific experimental constructions so far.

How to design an experimental protocol to observe this novel effect
in future experiments? To our knowledge, this effect has not been
explicitly and simply spelled out in any real physical system, which
is what we attempt to do here. In order to generalize to the
non-Abelian junction in experiments, we need a system of
multi-component order parameter which has a non-Abelian symmetry in
the order parameter space. In contrast with magnetic tarp, the spin
of the alkali atoms are essentially free in an optical trap
\cite{Stamper,Ohmi,Ho}. This spinor nature properly provide the
scenario of our non-Ablian construction. In the following, we
briefly introduce the system about a spinor atomic BEC in a
double-well optical potential. Although the dynamical tunneling
properties of spin-1 and ``pseudo spin-1/2" bonsonic systems were
calculated \cite{Ashhab,Ng,Leggett,L. You,L. You2}, the essence of
the non-Abelian effect has not been captured yet. In present Letter,
we focus on the pseudo Goldstone modes due to the non-Abelian
symmetry breaking, which is at the heart of Josephson effect. For
concrete construction, we propose spin-2 BEC in double optical trap.
The spin-2 system has possible advantages, compared to the spin-1
system, in the sense that the symmetry properties are much richer to
explore non-Abelian effects.

\emph{Ground state structure}.--- Let us consider a system of a
homogenous spin-2 Bose gas with s-wave interaction. This system can
be described by the following mean field free energy
\begin{eqnarray}
F(\psi)=\frac{1}{2}[c_{0}(\psi^{\dag}\psi)^{2}+c_{1}\langle\mathbf{f}\rangle^{2}
+\frac{c_{2}}{5}|\Theta|^2]-\mu\psi^{\dag}\psi,
\end{eqnarray}
where $\psi=(\psi_{2},\psi_{1},\psi_{0},\psi_{-1},\psi_{-2})^{T}$ is
the order parameter of the spin-2 Bose system, $c_{0}, c_{1}$ and
$c_{2}$ are interaction strengths related to the scattering length
in different spin channels,
$\langle\mathbf{f}\rangle=\psi^{\dag}\mathbf{f}\psi$ is the mean
value of the spin operator and
$\Theta=\sum_{a=-2}^{2}(-1)^{a}\psi_{a}\psi_{-a}$ represents a
single pair of identical spin-2 particles. The ground state
configuration can be determined by minimizing this free energy.
There are several distinct phases in this system
\cite{Ciobanu,Ueda}. We will discuss these phases under zero
magnetic field and analyze the symmetry and low-lying excitation
spectrum of each phase.

(I) Ferromagnetic phases: When $c_{1}<0$ and $c_{1}-c_{2}/20<0$, two
kinds of ferromagnetic phases are energetically favored. The
corresponding ground state configurations are given by
$\psi=\sqrt{n}e^{i\theta}(1,0,0,0,0)$ or
$\psi=\sqrt{n}e^{i\theta}(0,1,0,0,0)$, where $n=\mu/(c_{0}+4c_{1})$
is the particle density and $\theta$ is an arbitrary global phase.
It is obvious that these ground states have a $U(1)$ symmetry which
leads to only one massless Goldstone mode. Therefore, two uncoupled
system have a $U(1)\bigotimes U(1)$ symmetry. This symmetry will
soft breaks into a $U(1)$ diagonal symmetry when a weak coupling is
applied. This pattern of symmetry breaking corresponds to an Abelian
Josephson effect. The low-lying excitation spectrum of this state
has been derived as $
\omega_{k}=\sqrt{\epsilon_{k}(\epsilon_{k}+2g_{4}n)}$ \cite{Ueda},
where $\epsilon_{k}=k^{2}$ and $g_{4}=c_{0}+4c_{1}$. We should note
that this Goldstone mode will break into two modes when coupling is
applied: one zero energy mode and one pseudo Goldstone mode. This
pseudo Goldstone mode has a finite but small gap and leads to a
density mode fluctuation in d.c. Josephson current.

(II) Antiferromagnetic phase: In the absence of magnetic field,
there is only one kind of antiferromagnetic phase when $c_{2}<0$ and
$c_{1}-c_{2}/20>0$ are satisfied. The corresponding ground  state
configuration is degenerate with respect to five continuous
variables \cite{Ciobanu}. According to Goldstone's theorem, there
always exist massless modes in the ordered state when a continuous
global symmetry is spontaneously broken. As a result, the above
degeneracy of ground state will lead to five massless Goldstone
mode. Some of these modes will lead to non-Abelian Josephson effect
because the fluctuation of different components of the spinor BEC in
the each trap are coupled in the corresponding equations of motion.
Similar to the ferromagnetic case, we will show that each of these
modes leads to a pseudo Goldstone mode. These pseudo Goldstone modes
will give rise to five different kinds of Josephson currents with
different frequencies.

(III) Cyclic phase: When $c_{1}>0$ and $c_{2}>0$, the cyclic phase
is energetically favored. The ground state configuration is given by
$\psi=\sqrt{n}(\frac{e^{i\theta_{2}}}{2},0,\frac{e^{i\theta_{0}}}{\sqrt{2}},0,
\frac{e^{i\theta_{-2}}}{2}) $ where $n=\mu/c_{0}$ is the particle
density and the global phase $\theta_{\pm2}$ and $\theta_{0}$
satisfy $\theta_{2}+\theta_{-2}-2\theta_{0}=\pi$ \cite{Ciobanu}. We
find that there are two independent continuous phase variables in
the cyclic ground state. As a result, there should be at least two
Goldstone modes. We will show that these modes also lead to
non-Abelian Josephson effect.

\emph{Non-Ablian Josephson effects}.--- We will analyze the
Josephson effect of a spin-2 BEC system in a double-well optical
trap, as shown in Fig. 1. We assume that the energy barrier between
the two wells is strong enough so that the coupling between the Bose
gas in each well is very weak and the overlap of the ground state
wave functions in left and right well (which we denote as
$\varphi_{L}(\textbf{x})$ and $\varphi_{R}(\textbf{x})$) can be
safely neglected. We will also use the single mode approximation
which means we take the same mode function for all five spin
components, this is a widely used approximation and it is valid when
the spin interaction is symmetric. Under these assumptions, the
system can be described by the following potential
\begin{eqnarray}
V_{couple}=V(\psi_{L})+V(\psi_{R})-J(\psi_{L}^{\dag}\psi_{R}+\psi_{R}^{\dag}\psi_{L}),
\end{eqnarray}
where $\psi_{L}$ and $\psi_{R}$ are the order parameter of the Bose
system in left and right well respectively, $J=-\int
d^{3}\textbf{x}\varphi_{L}^{*}(\textbf{x})(-\nabla^{2}+V_{well}(\textbf{x}))\varphi_{R}(\textbf{x})$
is the coupling parameter, and $V(\psi)=F(\psi)$ has been defined in
equation (1) with a redefinition of the following parameters:
$c_{i}$ is redefined as $ c_{i}\int
d^{3}\textbf{x}|\varphi_{L}(\textbf{x})|^{4}$, $\mu$ is redefined as
$\mu=-\int
d^{3}\textbf{x}\varphi_{L}^{*}(\textbf{x})(-\nabla^{2}+V_{well}(\textbf{x}))\varphi_{L}(\textbf{x})=-\int
d^{3}\textbf{x}\varphi_{R}^{*}(\textbf{x})(-\nabla^{2}+V_{well}(\textbf{x}))\varphi_{R}(\textbf{x})$
which is just the single particle energy in each well. It should be
noted that we have taken the same chemical potential for the Bose
gas in right and left well, because we will only be interested in
the d.c. Josephson effect which captures the essence of the
non-Abelian symmetry breaking as simple as possible. With this
potential, we can derive the equation of motion of the Josephson
current in each phase and analyze the relationship between the
coupling parameter and the pseudo goldstone modes. However, as we
have mentioned above, the symmetry of the ground state in
ferromagnetic phase only leads to an Ablian Josephson effect which
is not interested in present paper. Therefore, we will just analyze
the antiferromagnetic phase and cyclic phase which are important
realizations of non-Ablian Josephson effect.

\begin{figure}[t]
\includegraphics[bb=0 0 506 380
height=2.0523in, width=2.6194in ]{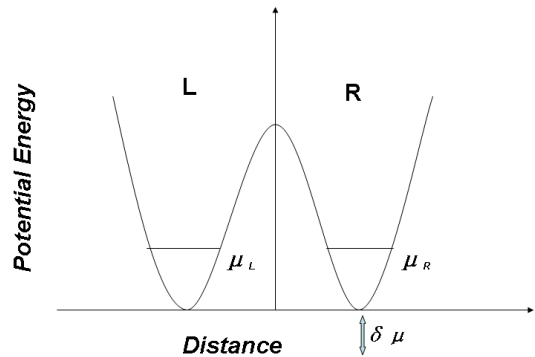}
\caption{\label{fig:epsart}The experimental schematic of a spin-2
Bose gas trapped in a double well with chemical potentials $\mu_{L}$
of left and $\mu_{R}$ of right trap, which initially satisfy
$\mu_{L}=\mu_{R}$ . To drive Josephson effect, we add a small
distortion $\delta\mu$ to $\mu_{R}$.}
\end{figure}

(I) Antiferromagnetic phase: To obtain the equation of motion of the
Josephson current, we can add a small fluctuation around the ground
state $\psi_{L}=\psi_{R}=\frac{\sqrt{n}}{\sqrt{2}}(1,0,0,0,1)$ and
we should first determine the ground state particle density $n$
\cite{Ho}. By minimizing the potential $V_{couple}$ under the above
antiferromagnetic ground state, we get $n=(\mu+J)/(c_{0}+c_{2}/5)$.
With this ground state configuration, we can obtain the equations of
the fluctuations in this phase and analyze the excitation spectrum.

The $m=0$ mode. The equation of motion of $m=0$ mode is given as
\begin{eqnarray}
i\frac{d}{dt}\phi_{0L}=(-\frac{c_{2}}{5}n+J)\phi_{0L}+\frac{c_{2}}{5}n\phi_{0L}^{*}-J\phi_{0R}\label{a0L},
\end{eqnarray}
and a similar set of equations of $\phi_{0R}$. Since this mode is
decoupled from others, it corresponds to an Abelian Josephson
current. By solving equation (\ref{a0L}), we obtain the
eigenenergies of this mode, one is zero corresponding to a massless
Goldstone mode and the other is
$\omega_{0}=2\sqrt{J(J+\frac{2|c_{2}|n}{5})}$ corresponding to a
pseudo Goldstone mode.

The $m=\pm1$ coupled mode. The $m=\pm1$ mode are coupled in the
following equations
\begin{eqnarray}
i\frac{d}{dt}\phi_{1L}&=&[n(c_{1}-\frac{c_{2}}{5})+J]\phi_{1L}\nonumber\\&+&n(c_{1}
-\frac{c_{2}}{5})\phi_{-1L}^{*}-J\phi_{1R},
\nonumber\\
i\frac{d}{dt}\phi_{-1L}&=&[n(c_{1}-\frac{c_{2}}{5})+J]\phi_{-1L}\nonumber\\&+&n(c_{1}
-\frac{c_{2}}{5})\phi_{1L}^{*}-J\phi_{-1R},
\end{eqnarray}
and a similar set of equations of $\phi_{\pm1R}$. The solution
involves one zero energy oscillation and one pseudo Goldstone mode
with a gap of
$\omega_{\pm1}=2\sqrt{n(c_{1}-\frac{c_{2}}{5})J+J^{2}}$.

The $m=\pm2$ coupled mode. The $m=\pm2$ mode are coupled in the
following equations
\begin{eqnarray}
i\frac{d}{dt}\phi_{2L}&\!=\!&(\frac{ng_{4}}{2}\!+\!J)\phi_{2L}\!+\!\frac{ng_{4}}{2}\phi_{2L}^{*}-J\phi_{2R}\nonumber\\
&\!+\!&\frac{1}{2}n(c_{0}-4c_{1}\!+\!\frac{2}{5}c_{2})(\phi_{-2L}\!+\!\phi_{-2L}^{*}),
\nonumber\\
i\frac{d}{dt}\phi_{-2L}&\!=\!&(\frac{ng_{4}}{2}\!+\!J)\phi_{-2L}\!+\!\frac{ng_{4}}{2}\phi_{-2L}^{*}-J\phi_{-2R}\nonumber\\
&\!+\!&\frac{1}{2}n(c_{0}-4c_{1}\!+\!\frac{2}{5}c_{2})(\phi_{2L}\!+\!\phi_{2L}^{*}),
\end{eqnarray}
and a similar set of equations of $\phi_{\pm2R}$. Through this set
of equations, we obtain two zero energy modes and two pseudo
Goldstone modes with energy gap
$\omega_{\pm2}^{(1)}=2\sqrt{n(c_{0}+\frac{c_{2}}{5})J+J^{2}}$ and
$\omega_{\pm2}^{(2)}=2\sqrt{n(c_{1}-\frac{c_{2}}{20})J+J^{2}}$. In
Fig. 2 we give the dependence of the above frequencies on coupling
parameter $J$. Recently, a polar behavior has been observed in the
F=2 ground state of $^{87}Rb$ condensate \cite{Erhard}. We expect
that the above modes of fluctuations can be observed in this system
in future experiments. In the case of $^{87}Rb$ system, the value of
interacting strengths under typical experimental condition are given
as \cite{Erhard}: $c_{1}n:0-10$nK, $c_{2}n:0-0.2$nK and $c_{0}n$
about 150nK. According to the weak coupling limit, we assume that
the coupling parameter $J$ is much smaller than the interaction
energy of the condensate and given as about $0.1$nK. Under these
conditions, we can obtain the frequencies of the fluctuation related
to the antiferromagnetic phase, which is of order $100$Hz. The
measurement of fluctuations on this characteristic time scale (about
10ms) is accessible in current experiments.

\begin{figure}[t]
\includegraphics[bb=156 269 439 548
height=2.85900in, width=2.96000in ]{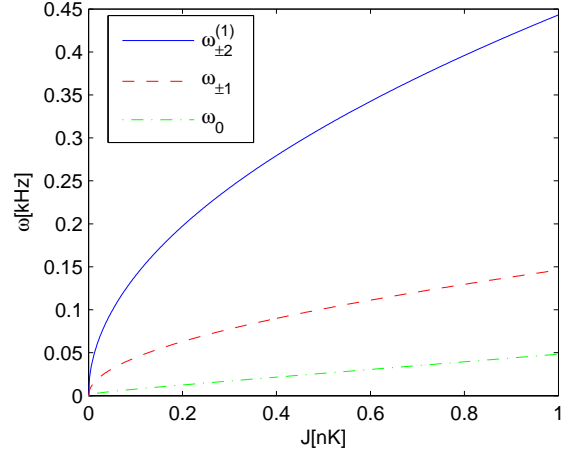}
\caption{\label{fig:epsart}The frequencies of the pseudo Goldstone
modes as a function of coupling parameter $J$ in the case of
antiferromegnetic phase. All the frequencies is proportional to
$\sqrt{J}$ when $J$ approaches zero and the dependence on $J$
becomes linearizing when $J$ is large compared to the interaction
energy.}
\end{figure}

(II) Cyclic phase: Following the same procedure in the
antiferromagnetic phase we can get the equations of motion for the
cyclic phase.

The $m=\pm2,0$ coupled mode. The fluctuations of $m=\pm2$ and $m=0$
mode are coupled together and can be described by the following
equations
\begin{eqnarray}
i\frac{d}{dt}\phi_{2L}&\!=\!&(\frac{\alpha\!+\!\beta\!+\!\gamma}{4}\!+\!J)\phi_{2L}
\!+\!\frac{\alpha\!+\!\beta}{4}\phi_{2L}^{*}\nonumber\\&\!+\!&\frac{\alpha\!-\!\beta}{4}\phi_{-2L}^{*}
\!+\!\frac{1}{4}(\alpha\!-\!\beta\!+\!\gamma)\phi_{-2L}\!+\!\frac{\alpha}{2\sqrt{2}}\phi_{0L}^{*}
\nonumber\\&\!+\!&\frac{\alpha-\gamma}{2\sqrt{2}}\phi_{0L}-J\phi_{2R},
\nonumber\\
i\frac{d}{dt}\phi_{0L}&\!=\!&(\frac{\alpha\!+\!\gamma}{2}\!+\!J)\phi_{0L}
\!+\!\frac{\alpha-\gamma}{2\sqrt{2}}(\phi_{2L}\!+\!\phi_{-2L})\nonumber\\
&\!+\!&\frac{\alpha}{2}\phi_{0L}^{*}\!+\!\frac{\alpha}{2\sqrt{2}}(\phi_{2L}^{*}\!+\!\phi_{-2L}^{*})-J\phi_{0R},
\nonumber\\
i\frac{d}{dt}\phi_{-2L}&\!=\!&(\frac{\alpha\!+\!\beta\!+\!\gamma}{4}\!+\!J)\phi_{-2L}
\!+\!\frac{\alpha\!+\!\beta}{4}\phi_{-2L}^{*}\nonumber\\&\!+\!&\frac{\alpha\!-\!\beta}{4}\phi_{2L}^{*}
\!+\!\frac{1}{4}(\alpha\!-\!\beta\!+\!\gamma)\phi_{2L}\!+\!\frac{\alpha}{2\sqrt{2}}\phi_{0L}^{*}
\nonumber\\&\!+\!&\frac{\alpha-\gamma}{2\sqrt{2}}\phi_{0L}-J\phi_{-2R},
\end{eqnarray}
and a similar set of equations of $\phi_{\pm2R}$ and $\phi_{0R}$.
The parameters in the above equations are defined following,
$\alpha=c_{0}n,\beta=4c_{1}n$ and $\gamma=2c_{2}n/5$. By solving the
above equations, we find that each Goldstone mode in the
corresponding uncoupled system \cite{Ueda} break into one massless
mode and one pseudo Goldstone mode. Since there are two Goldstone
modes in the uncoupled system, we find two pseudo Goldstone modes
with energy $\omega_{0,\pm2}^{(1)}=2\sqrt{J^2+\alpha J}$ and
$\omega_{0,\pm2}^{(2)}=2\sqrt{J^2+\frac{\beta}{2} J}$.

The $m=\pm1$ coupled mode. The fluctuations of $m=\pm1$ mode are
also coupled together and can be described by the following
equations
\begin{eqnarray}
i\frac{d}{dt}\phi_{1L}\!=\!(\frac{\beta}{2}\!+\!J)\phi_{1L}\!+\!\frac{\beta}{4}\phi_{-1L}^{*}
\!+\!\frac{\sqrt{3}\beta}{4}\phi_{1L}^{*}\!-\!J\phi_{1R},\nonumber\\
i\frac{d}{dt}\phi_{-1L}\!=\!(\frac{\beta}{2}\!+\!J)\phi_{-1L}\!+\!\frac{\beta}{4}\phi_{1L}^{*}
-\frac{\sqrt{3}\beta}{4}\phi_{-1L}^{*}\!-\!J\phi_{-1R},
\end{eqnarray}
and a similar set of equations of $\phi_{\pm1R}$. As we know, there
are two massless Goldstone modes with the same energy in the
uncoupled system \cite{Ueda}. By solving the above equation of
motion, we can see that each of them leads to a pseudo Goldstone
mode with a gap $\omega_{\pm1}=2\sqrt{J^2+\frac{\beta}{2} J}$. These
kind of fluctuations in cyclic phase is expected to be realized in a
condensate of $^{85}Rb$ atoms \cite{Ciobanu}. Based on the current
estimates of scattering lengths, the value of interacting strengths
under typical experimental condition are given as: $c_{1}n:0-20$nK,
$c_{2}n:0-0.6$nK and $c_{0}n$ about 600nK. Under these conditions,
we can also estimate the d.c. Josephson frequencies in cyclic phase
which is about 100-300Hz. The dependence of the above frequencies on
coupling parameter $J$ is shown in Fig.3.

All the above analysis is base on the assumption that the phase of
the condensates in two wells are equal to each other in their ground
state. However, there is another experimental achievable
configuration called $\pi$ state in which the phases of condensates
in right and left well have a $\pi$ difference. We find that,
analogue to spin-1 and single component BEC system
\cite{Ashhab,Niu,Jin}, the oscillation around this state could be
unstable in some cases. For example, in the case of
Antiferromagnetic phase, the value of pseudo Goldstone mode
$\omega_{0}$ will change to
$\omega_{0}=2\sqrt{J(J-\frac{2|c_{2}|n}{5})}$ in $\pi$ state. When
$J<\frac{2|c_{2}|n}{5}$, this mode can clearly become unstable. This
in fact happens in all the pseudo Goldstone modes we calculated
above. As a result, the experimental protocol we propose below
should only restricted in the oscillation around the true ground
state of the system not around the $\pi$ state.

\begin{figure}[t]
\includegraphics[bb=166 289 429 521
height=2.31000in, width=2.42000in ]{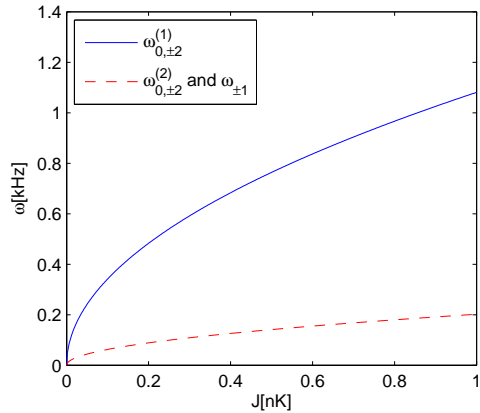}
\caption{\label{fig:epsart}The frequencies of pseudo Goldstone modes
as a function of coupling parameter $J$ in the case of cyclic phase.
The dependence on $J$ is similar to the case in antiferromegnetic
phase. However, there are two degenerate modes
$\omega_{0,\pm2}^{(2)}$ and $\omega_{\pm1}$ in this case, which are
originated from the two degenerate Bogoliubov modes in the uncoupled
system.}
\end{figure}

\emph{Experimental signatures of Non-Abelian Joesphson effects}.---
We want to propose an experimental protocol to observe the
non-Abelian Josephson effect. The experimental set up of a spin-2
Bose gas trapped in a double well is illustrated in Fig. 1. The d.c.
non-Abelian Josephson current can be detected with the following
steps. The first step is to initiate a density oscillation in the
system. This can be realized by slightly changing the depth of one
well, which will cause a small imbalance in chemical potential
($\mu_{R}\rightarrow \mu_{R}+\delta\mu$) between the two wells, and
then tune it back. The next step is to detect the time dependence of
the particle numbers in different spin component. Such kind of
detection can be realized by first spatially separating different
spin component with a Stern-Gerlach method during time of flight
after switching off the trapping potential. Then, the number of
atoms in each spin component will be related to the respective
spatial density distributions which can be evaluated by the
absorption imaging method. Following the above steps, one can
measure the density oscillation in each spin component which are
coupled together due to the non-Abelian symmetry of the system. The
measurement in the dependence of the oscillation frequencies on $J$
can be realized by varying the barrier between the two wells and
repeating above measurement.

In mean field theory, the condensates of $^{87}Rb$ atom in F=2 state
is predicted to be polar $(c_{1}-c_{2}/20> 0$ and $c_{2} < 0)$, but
close to the border to the so-called cyclic phase $(c_{1} > 0$ and
$c_{2} > 0)$ \cite{Ciobanu}. Furthermore, polar behavior in the F=2
ground state of $^{87}Rb$ has been observed in recent experiment
\cite{Erhard}. As a result, we expect that the pseudo Goldstone
modes of the antiferromegnetic phase could be observed in
experiments. As we have mentioned, the value of interacting
strengths under typical experimental condition are given as
\cite{Erhard}: $c_{1}n:0-10$nK, $c_{2}n:0-0.2$nK and $c_{0}n$ about
150nK, which leads to the time scale of about 10ms in the
fluctuations in this system. On this time scale, the measurement we
proposed above is completely accessible in recent experiment in F=2
spinor Bose-Einstein condensates of $^{87}Rb$ system \cite{Erhard}.
Although there is still no such kind of measurement performed in a
system with cyclic phase, we expect that it will be realized in a
condensate of $^{85}Rb$ atoms in near future \cite{Ciobanu}.

In summary, we propose an experimental protocol to realize the so
called non-Abelian Josephson effect in a spinor Bose system. We find
that the frequencies of pseudo Goldstone modes do not only relate to
the coupling parameter but also to the interacting strengthes, which
is a nonlinear effect due to the spin dependent interaction. Our
results are of particular significance for exploring the new
features of the non-Abelian Josephson effects which is very
different from the Abelian case.

We thank Professors G. R. Jin, L. D. Carr and J. Brand for helpful
discussions. This work is supported by the NSF of China under Grant
Nos. 10874235, 60525417, 10740420252, the NKBRSF of China under
Grant 2005CB724508, 2006CB921400.

\end{document}